\documentstyle[aps]{revtex}

\title{\hfill \begin{small} Preprint: WSU--NP--98--2 \end{small}\\
${\bar \Lambda}/{\bar p}$ RATIOS IN HEAVY ION COLLISIONS AT 
$11.6~{\rm A\cdot GeV}/c$}
\author{\protect G.\,J.~Wang, G.~Welke, R.~Bellwied, and C.~Pruneau}

\address{Department of Physics and Astronomy\\ Wayne State University, 
Detroit, MI 48202, U.S.A.\\
E--mail: welke@physics.wayne.edu}

\begin{document}
\maketitle

\begin{abstract}
  We attempt to explain the ${\bar \Lambda}/\bar p$ ratios measured in
  heavy ion collisions at $11.6~{\rm A\cdot GeV}/c$ beam momentum
  within a hadronic framework.  This ratio is enhanced relative to
  corresponding ratios in $pp$ collisions, and is large when compared 
  with thermal fits to heavy ion data.  Using a detailed cascade
  calculation, we show that different annihilation cross--sections of
  $\bar \Lambda$'s and $\bar p$'s, and the net conversion of $\bar
  p$'s to $\bar \Lambda$'s, do not account for the enhancement in
  central collisions.  For larger impact parameters, however, hadronic
  mechanisms may well suffice to produce the observed enhancement.
  Uncertainties in elementary cross--sections and formation times are
  considered.
\end{abstract}

\section{INTRODUCTION}

Anti--lambda ($\bar \Lambda$) cross--sections in heavy ion collisions
are of interest because strangeness production is a potential signal
for the formation of a quark--gluon plasma (QGP) in heavy ion
collisions \cite{shuryak,koch,rafelski}.  The ratio of anti--lambdas
to anti--protons (${\bar p}$) is of particular interest since it
reflects the production of strange anti--quarks relative to
non--strange light anti--quarks, and should increase substantially
relative to the production expected from a superposition of $NN$
collisions, if a QGP is formed.

Recent experiments have reported measurements of $\bar p$ and $\bar
\Lambda$ production in various heavy ion systems at the Brookhaven AGS
and at the CERN SPS \cite{e859,e864a,e864b,na49,na35}.  Experiment E859
reports the ratio of $\bar \Lambda$ and $\bar p$ rapidity
distributions to be $3 \pm 1 \pm 1$ in central Si+Pb collisions
\cite{e859}. This ratio is corrected for finite experimental
acceptance, efficiencies, and for $\bar p$ creation from $\bar
\Lambda$ decay (feed-down), and also takes into account that neutral
$\bar \Sigma$ particles cannot be distinguished from the $\bar
\Lambda$ sample.  Experiment E864 has measured \cite{e864a,e864b} the $\bar
p$ production cross--section in Au+Pb at $11.6~{\rm A\cdot GeV}/c$.
They compare this measurement with a similar one from the E878
collaboration that was obtained with a focusing spectrometer, and
interpret the difference between the two measurements as an indicator
of $\bar \Lambda$ production.  E864 estimates that $\bar \Lambda/\bar
p > 2.3$ at the $98\%$ CL, with a most probable value of 3.5. At
the SPS, NA35 has published $\bar \Lambda/\bar p$ ratios for $pp$,
$pA$, S+S, S+Ag, and S+Au collisions at $200~{\rm A\cdot GeV}/c$, and
they observe a significant rise from 0.25 for $pp$-collisions to 1.5
for the heavy ion systems \cite{na49,na35}.

It is tempting to interpret the large reported ratios as evidence for
the formation of a QGP: Many authors \cite{letessier,let2,sollfrank}
have argued that the CERN multi-strange baryon ratios
\cite{judd,dibari,kinson} can only be described by a QGP scenario.
Moreover, in a systematic study of strangeness production in heavy ion
collisions from SIS to SPS energy ranges, mostly based on kaon ratios,
Geiss et al. \cite{Geiss} conclude that a QGP scenario is required to
explain that data at AGS energies.  This conclusion is, however,
challenged by Au+Au RQMD simulations that find that the $\bar p$
production cross--section appears to be lower than expected from the
scaled $NN$ $\bar p$ cross--section.

Here we shall try to address quantitatively all possible hadronic
contributions to the $\bar \Lambda/\bar p$ ratio using a cascade model
\cite{WW}.  We restrict calculations to AGS energies and show that
``differential annihilation'' of the two species and $\bar
p$--to--$\bar \Lambda$ conversion processes can indeed enhance the
$\bar \Lambda/\bar p$ ratio. We conclude that the effect is not large
enough at central impact parameters, but may well suffice to explain
the data from more peripheral events -- thus hinting at a production
mechanism outside of standard hadronic interactions.

We present our arguments as follows: Section II outlines a thermal
model calculation with chemical equilibration and feed-down
contributions. We fit the parameters of the model to rather liberal
ranges of various kaon, pion and proton ratios measured in central
Au+Au events, and show that the resulting $\bar \Lambda/\bar p$ ratio
lies well below the E864 value. This conclusion has also been reached
by other authors \cite{pbm1,pbm2,heinz}.  We consider transport
simulations in the remainder of the paper, briefly describing the model
\cite{WW} and discussing the relevant cross--sections in more detail.
Actual measurements of the $\bar \Lambda$ and $\bar p$ cross--sections
\cite{gjesdal,eisele} are used, rather than event generator
parameterizations. Section~IV presents our results, discussing also the
uncertainties in the calculations, particularly those arising from the
relatively poorly known $\bar \Lambda$ annihilation in the relevant
energy range, and from particle formation times.  We shall consider
first central events, and then the more recent E864 results
\cite{QM97E864} at larger impact parameters.  Speculative conclusions
based on the quantitative discrepancy between calculations and the
actual measurements at small impact parameters are presented in the
last Section.

\section{THERMAL MODEL}

We consider in this Section a thermal model fit for Au+Au collisions at
$11.6~{\rm GeV}/c$, at central impact parameters only.  Our main
motivation is to show that in such events the measured ${\bar
  \Lambda}/{\bar p}$ ratio is larger than predicted by a standard
thermal model, even if we push the input fit ratios to extreme values.

The model is based on thermal and chemical equilibration 
up to some freeze--out, so that
all relative abundances can be obtained from four parameters: the
freeze--out temperature $T_0$, the electric chemical potential $\mu_e$,
the baryonic chemical potential $\mu_b$, and the strange chemical
potential $\mu_s$.  Here,
\begin{eqnarray}
 \mu_i \;=\; q_i\,\mu_e \:+\: b_i\,\mu_b \:+\: S_i \, \mu_s \nonumber 
\end{eqnarray}
are particle chemical potentials, where $q_i$, $b_i$ and $S_i$ are the
charge, baryon number and strangeness of species $i$, respectively.
Note that the strange chemical potential is not fixed {\it a priori}.
The distribution functions we use are for ideal Fermions and Bosons --
we do not include mean fields or excluded volume modifications
\cite{Cleymans}, which would alter our conclusions quantitatively, but
not qualitatively.  Of course, the final abundance of a particle is
determined by both the primary number of this species at freeze--out,
and by feed--down from heavier species after freeze--out.  We include
all mesons with rest mass $\le 1~{\rm GeV}/c^2$, and all baryons with
rest mass $\le 1.7~{\rm GeV}/c^2$. We note that our final $\bar
\Lambda$ numbers throughout include ${\bar \Sigma}^0$'s, as none of the 
experimental measurements can distinguish between the two species.

The results of the calculation are summarized in Table~1.  The thermal
fit parameters $T_0, \mu_e, \mu_b$ and $\mu_s$ are obtained by
applying the conditions $|Q/B-0.40| < 1\%$ and $|S|< 2\%$, where $Q$,
$B$ and $S$ are the overall charge, baryon number and strangeness of
the system, together with the constraints listed in Table~1.  As
motivated above, we purposefully choose rather large ranges in these
latter constraints to show how difficult it would be to obtain the
experimental ${\bar \Lambda}/{\bar p}$ ratio.  The ``errors'' on the
fitted values for $T_0$ and the chemical potentials listed in the
caption of Table~1 indicate roughly the ranges that lead to results
consistent with the constraint intervals.\footnote{We note that these
  parameters are not inconsistent with overall energy conservation
  considerations.}

While the agreement of most ratios with the data is relatively good
(see Table~1, also Refs.~\cite{pbm1,pbm2}), we see that ${\bar
  \Lambda}/{\bar p}\, {\buildrel < \over \sim}\, 1.9$, at best, well
below the 98\% CL lower value of 2.3 reported by E864.  In Figure~1
these results are illustrated graphically.  Panel (a) shows ${\bar
  \Lambda}/{\bar p}$ as a function of the $K^{+}/K^{-}$ ratio for
various freeze--out temperatures $T_0$.  Large ${\bar \Lambda}/{\bar
  p}>2$ result only if the freeze--out temperature and/or observed
$K^+/K^-$ ratio are pushed unreasonably high.  A similar conclusion
follows from Figure~1(b), which shows ${\bar \Lambda}/{\bar p}$ as a
function of $K^+/\pi^+$. In both panels the error bars/scatter points
indicate values consistent with constraints not shown, and the dashed
line represents the 98\% lower CL of the E864 experiment.

Clearly, the experimental ${\bar \Lambda}/{\bar p}$ ratio has, at the
very least, a significant non--thermal component: the lower
experimental bound reported by E864 is larger than any reasonable
thermal fit would allow.

\section{CASCADE MODEL AND CROSS--SECTIONS}

Generally, a thermal description is useful for a given particle
species if its mean free path is small compared to the system size.
This fact alone means that other approaches should also be
investigated. We do so in the next Section, and discuss here the
relevant cross--sections.

Our cascade model will be described in detail elsewhere \cite{WW}, and
we give here only a brief overview.  It is used in this work to
describe the evolution of the pions, kaons, and nucleons which act as
a ``background'' for the ${\bar p}$ and ${\bar \Lambda}$ species that
interest us.  The model is based on $NN$ input data, and includes
elastic and inelastic processes.  For elastic collisions, resonance
formation is considered if the energy and quantum numbers of the
combined system are suitable. For inelastic collisions, we consider
single diffractive processes, which effectively correspond to
resonance production at low energies, and non--single diffractive
processes, responsible for the remainder of the inelastic
cross--section.  All cross--sections are taken from experimental data
whenever available, otherwise we use $NN$, $\pi N$ or $KN$, and
$\pi\pi$ or $\pi K$ cross--sections for baryon--baryon, meson--baryon,
and meson--meson interactions, respectively.  Average particle
numbers, the multiplicity distribution, and species abundances are
also taken from data whenever available.  For leading particles, the
fractional longitudinal momentum is determined from a uniform
distribution, while the transverse momentum is taken from 
$p_{t}\exp(-\beta p_t)$ with $\beta\approx 4.44~{\rm (GeV/c)^{-1}}$.  The
transverse momenta of secondary particles are determined in a similar
way, while their rapidities are chosen from a flat distribution. All
conservation laws are strictly enforced.

The following processes are of particular importance for our discussion.
Firstly, the production of $\bar p$'s and $\bar \Lambda$'s:
\begin{eqnarray}
b_{1}+b_{2}&\rightarrow& N_{1}+N_{2}+B+\bar B~(+\pi)
\label{production_f}\\
b+M&\rightarrow& N+B+\bar B~(+\pi)
\label{production_2}\\
M+M&\rightarrow& B+\bar B~(+\pi)
\label{production_3}\\
S^{+}+S^{-}&\rightarrow& B+\bar B~(+\pi)
\label{production_4}\\
b+S^{+}&\rightarrow& b+{\bar B}+N~(+{K^+}+\pi)
\label{production_5}\\
M+S^{+}&\rightarrow& N+{\bar B}~(+S^{+}+\pi) \label{production_l}
\end{eqnarray}
Here, $b$ represents a non-strange baryon, $N$ a nucleon, $B$ any
baryon, $M$ a light unflavored meson, and $S^{\pm}$ a $\pm
1$--strangeness meson. We return to the question of production below.
Secondly, in the same notation:
\begin{eqnarray}
b+\bar{p}&\rightarrow& X \label{annihilation_f}\\
b+\bar{\Lambda}&\rightarrow& S^{+}+\pi \label{annihilation_2}\\
\bar{b}+S^{+}&\rightarrow&{\bar \Lambda}+\pi\label{annihilation_3}\\
\bar{p}+M&\rightarrow&{\bar \Lambda}+\pi+S^{-} \label{annihilation_4}\\
\bar{\Lambda}+M&\rightarrow&{\bar b}+S^{+}(+\pi)\label{annihilation_l}
\end{eqnarray}
We refer to (\ref{annihilation_f}) and (\ref{annihilation_2}) as
annihilation processes, and
(\ref{annihilation_3})--(\ref{annihilation_l}) as conversion
processes. The inverse for all $2\rightarrow 2$ body processes listed
is also included, but we have not considered $3,4,\ldots 
\rightarrow 2$ detailed balance.

The experimental $p\bar{p}$ annihilation cross--section is well known,
and can be parameterized as
\begin{eqnarray}
\sigma_{p\bar{p}}^{annih}(p_{lab})\;=\;67\:p_{lab}^{-0.7}~{\rm mb}~,
\label{PPbar}
\end{eqnarray}
where $p_{lab}$ is the momentum of the ``beam'' particle in ${\rm
  GeV}/c$ with the ``target'' at rest. The solid line in Figure~2
shows this parameterization of the data (crosses).  The
$\bar{\Lambda}$ annihilation cross--section, on the other hand, is
relatively poorly known, especially in the energy range we are
interested in. We model it by assuming that the elastic
cross--sections for $p\bar{\Lambda}$ and $p\Lambda$ are equal, and
then use the data of Ref.~\cite{eisele} to obtain:
\begin{eqnarray}
\sigma_{p\bar{\Lambda}}^{annih}(p_{lab})\;=\;15\:
(p_{lab}/10)^{-\alpha}~{\rm mb}~, \label{PLbar_low}
\end{eqnarray}
where the same comments apply as for Eq.~(\ref{PPbar}).  While the
data is best fit by $\alpha=0.5$, the uncertainty is rather large (see
Figure~2; the diamonds are data from Ref.~\cite{eisele}).  In fact,
one might well argue that the ${\bar \Lambda}$ data is consistent with
$\alpha=0$, as shown by the dot--dashed line in Figure~2. We take
$\alpha=0$ as an extreme value that should lead to the greatest
differential annihilation.  For $\alpha=0.7$, on the other hand, the
$\bar \Lambda$ data is practically indistinguishable from the $\bar p$
data.  We shall subsequently investigate the behavior of $\bar
\Lambda/\bar p$ with $\alpha$.

The $\bar{p}$ and $\bar{\Lambda}$ scattering with mesons is also an
important process we need to consider. Broadly speaking, we have three
types of collision: (1) thermalization of $\bar{p}$'s and
$\bar{\Lambda}$'s through elastic collisions; (2) production of
resonances that eventually decay back into $\bar{p}$'s or
$\bar{\Lambda}$'s; and (3), most importantly, net conversion of $\bar
p$'s to $\bar{\Lambda}$'s.  Chief amongst these is
\begin{equation}
\bar{p}+K^{+}\rightarrow \pi+\bar{\Lambda}~ ~ ~ ({\mbox {or resonances of }} 
\bar{\Lambda})\label{PKp}~ ~ ~,
\end{equation}
for which we know the charge conjugate reaction to have a sizeable
cross--section. The process (\ref{PKp}) thus contributes to reducing
the $\bar{p}$ abundance while enhancing the $\bar{\Lambda}$ abundance
in the final state.  Given the pronounced strangeness enhancement in
large systems such as Au+Au, the process (\ref{PKp}) should be a
relatively important piece of the $\bar{\Lambda}/\bar p$ ``puzzle.''

The production rate of $\bar p$'s or $\bar{\Lambda}$'s is only
$O(10^{-2})$ per event at the AGS, and cascade calculations are rather
CPU intensive. We shall therefore perform an effective calculation of
the survival probability by putting $\bar p$'s or $\bar{\Lambda}$'s in
by hand, wherever and whenever a collision occurs in which the energy
is sufficient for a $p\bar p$ or $\Lambda\bar{\Lambda}$ pair to be
produced. Thus we assume that the pair production does not depend on
energy, once above threshold. and take the relative initial production
of ${\bar p}$'s and ${\bar \Lambda}$'s from $pp$ data (see below).
The presence of initial $\bar p$'s or $\bar{\Lambda}$'s is not
permitted to influence the evolution of nucleons, pions and kaons --
we restore particles that interacted with $\bar p$'s or
$\bar{\Lambda}$'s to their pre--collision kinematics. 

We may then calculate the ${\bar \Lambda}/{\bar p}$ ratio via the
survival probabilities $P_s$ and the number of converted anti--lambdas
that survive, ${\bar \Lambda}^c$:
\begin{equation} 
\frac{\bar \Lambda}{\bar p}\;\approx\;r_p\,\frac{P_s({\bar \Lambda})}
{P_s({\bar p})}\:+\:2 r_c\,\frac{{\bar \Lambda}^c}{{\bar p}^s}~ ~,
\label{final_ratio}
\end{equation} 
where ${\bar p}^s$ is the number of surviving ${\bar p}$'s.  Here,
$r_c\approx 0.25$ is a correction factor, as we overestimate $\bar
\Lambda$--production by assuming that all collisions of non-strange
anti--baryons with positive strange mesons result in a ${\bar
  \Lambda}$. The factor two accounts for the conversion of $\bar n$'s
to ${\bar \Lambda}$'s.  Finally, $r_p$ is the ${\bar \Lambda}/{\bar
  p}$ ratio in $pp$ collisions. At $\sqrt{s}\sim 20~{\rm GeV}$, it has
a value \cite{blobel,antinuc,whit,GadRoh} of 0.25--0.30. At AGS
energies ($\sqrt{s}\sim 5~{\rm GeV}$) its value is less established.
Using Refs.~\cite{blobel,antinuc,whit}, we infer a value of $\sim
0.2$, but we shall use a value of $r_p \approx 0.25 +0.0 -\!\!0.1$
throughout this work.  We note that mean field effects may very well
reduce the effective $r_p$ for in--medium production when compared
\cite{Cassing} to the free space values.  Our chosen value of $r_p$
and error bars imply that the heavy ion ${\bar \Lambda}/{\bar p}$
ratios we obtain in subsequent calculations will be upper bounds.

\section{CASCADE CALCULATION: RESULTS}

We begin by discussing the most central events, roughly $10\%$ of the
total cross--section.  To illustrate the effect of differential
annihilation versus net conversion, we show in Table~2 the different
components of Eq.~(\ref{final_ratio}) for different $\alpha$ and
$\tau=0$ (no formation time) in central Au+Au simulation at AGS
energies.  For small $\alpha$ (small ${\bar \Lambda}$ annihilation
cross--section), ${\bar \Lambda}/{\bar p}$ enhancement results almost
entirely from differential annihilation.  As $\alpha$ increases, the
survival probability of a ${\bar \Lambda}$ and a ${\bar p}$ become
equal, and the sole enhancement in the final ratio results from
conversion. Of course, as $\alpha$ increases the ratio 
${\bar \Lambda}^c/{\bar p}^s$ decreases, since ${\bar \Lambda}^c$ refers 
to surviving anti--lambdas.

We show in Figure~3 the final ${\bar\Lambda}/{\bar p}$ ratios from
cascade simulations (circles) for several values of $\alpha$ and
$\tau$, and for various systems, all at central impact parameters.
The dashed lines show the E864 98\% CL values for Au+Au. 
We clearly see that reducing the difference between ${\bar \Lambda}$
and ${\bar p}$ annihilation, {\it i.e.}, increasing $\alpha$, reduces
the ratio (Figure~3 (a)-(c)).  Similarly, a larger formation time
$\tau$ also leads to a smaller ratio -- the effective system size for
annihilation and conversion is reduced as all particles require some
time to be formed (see panel (d)). Moreover, for Au+Au, the maximum 
ratio obtainable (for $\alpha=0$ and no formation time) lies at about the 
the E864 98\% CL lower limit.

In discussing the results of Figure~3, we note that (1) realistic
values for $\alpha$ are probably closer to 0.5 than 0.0; (2) formation
times are probably not zero; and (3) our initial, {\it i.e.} $pp$
${\bar \Lambda}/{\bar p}$ ratio is probably an over--estimate,
particularly in light of possible medium effects that penalize ${\bar
  \Lambda}$ production relative to ${\bar p}$ production.  
This latter fact is reflected in the error bars for the ratio -- our choice
for $r_p=0.25 +\!0.0 -\!0.1$ is what we consider an upper bound. We conclude
that for central Au+Au events the most likely
value of ${\bar\Lambda}/{\bar p}$ is most definitely ${\buildrel <
  \over \sim}\, 2$, and more likely somewhat below unity -- smaller than
the 98\% CL reported by E864.

The ${\bar \Lambda}/{\bar p}$ ratio is larger in Si+Pb collisions than
in Au+Au collisions in our calculation, but it is not a large effect
(compare Figure~3 (a) and (c)).  Within our model this difference has a
simple geometric explanation: For Si+Pb, the $\bar p$'s and $\bar
\Lambda$'s are produced close to the beam axis, leading to a larger
average nuclear thickness that the anti--particle must traverse.
Once again, though, realistic input parameter values of $\tau\sim 1$ 
and $\alpha\sim 0.4$ seem to suggest that our hadronic cascade
misses the experimental lower bound of E859.

The solid lines in Figure~3 are results from a model calculation
\cite{GJWang} that uses annihilation and net conversion mean free
paths as an input to rate equations in a simplified geometric picture.
Its only free parameter mocks up the effect of local matter expansion,
local average momentum distributions, {\it etc.} It reproduces the
trends in the data rather well.  

Recently, E864 has also reported the impact parameter dependence of
the ${\bar\Lambda}/{\bar p}$ ratio in Au+Au \cite{QM97E864}. In
Figure~4, the upper and lower solid line (drawn to guide the eyes)
show the most probable (diamonds) and 98\% CL (squares) values of
these results, respectively. There are four centrality bins: From left
to right they are $<10\%$, 10-30\%, 30-70\%, and $>70\%$ of the total
cross--section, as obtained from transverse energy cuts. The
short--dashed line joins cascade calculations (crosses) for $\alpha=0$
and $\tau=0$.  Recall that both these parameters values are extreme:
$\alpha=0.4$ (box--crosses, joined by long dashes in Figure~4) is more
likely, and does not reproduce the central data bin result. For larger
impact parameters, however, we see that this conclusion is
considerably weakened -- the cascade result initially rises with
increasing impact parameter to far above the data for $\alpha=0$, while
for $\alpha=0.4$ it remains relatively flat, but eventually does
intercept the experimental data in semi--central collisions.

Thus, while hadronic mechanisms seem inadequate for the most central
events, we are not able to conclude with certainty that they fail to
describe the data at larger impact parameters.  We note that at higher
energies the difference between the ${\bar \Lambda}$ and ${\bar p}$
annihilation cross--section practically vanishes. Within the
differential annihilation model, one might therefore predict (in
hindsight) that ${\bar \Lambda}/{\bar p}$ at the SPS should be lower
than at the AGS, as indeed observed, and indicating that the model has 
the correct beam--energy dependence.

\section{CONCLUSIONS}

The ${\bar \Lambda}/{\bar p}$ ratio in $NN$ collisions at AGS energies
is ${\buildrel < \over \sim}\, 0.25$, heavy ion experiments measure
much larger values.  A hadronic thermal model with chemical
equilibration and feed--down contributions can only be pushed to
values above 2 with much difficulty.  For Au+Au, using the
(unreasonable) extremes of $K^{+}/\pi^{+}\sim 0.30$, $K^{+}/K^{-}\sim
6.0$, and $T_0\sim 140~{\rm MeV}$, we obtain an upper limit of ${\bar
  \Lambda}/{\bar p} \sim 2$, below the E864 98\%--lower confidence
limit of 2.3.

We have considered a non--equilibrium description, and presented
results from a detailed cascade calculation.  Chief inputs to our
calculations are the ${\bar \Lambda}$ and ${\bar p}$ annihilation
cross--sections with nucleons, as well as processes that convert
${\bar p}$'s into ${\bar \Lambda}$'s.  The ${\bar \Lambda}$
annihilation cross--section is relatively poorly known in the relevant
energy range, but it may be argued that it is somewhat less than
${\bar p}$ annihilation at AGS energies. Our parameter $\alpha$
controls this difference, with $\alpha=0$ corresponding to the
greatest difference, and $\alpha=0.7$ corresponding to a
cross--section that is practically indistinguishable from ${\bar p}$
annihilation. Currently, the cross--section data is (``best'') fit by
$\alpha=0.5$. Another important input is the cross section of $NN$ ${\bar
  \Lambda}$ to ${\bar p}$ production, which in our calculation may
well be an overestimate, compared to free $NN$ data and also in view of 
medium effects on the particle production. Our asymmetric error bars
on the input $NN$ ratio reflect this recognition.

In Au+Au cascade simulations for small impact parameters at AGS
energies, the largest ${\bar \Lambda}/{\bar p}$ ratio obtainable is
$\sim 2.4 +\!0.2 -\!1.0$, for a flat and small ${\bar \Lambda}$
annihilation cross--section ($\alpha=0$) and no formation time
($\tau=0$). Both these input values are extreme. What we currently
believe to be more reasonable inputs lead to a value of ${\bar
  \Lambda}/{\bar p} {\buildrel < \over \sim} 1$ ($\sim 1$ for Si+Pb),
far below the lower bound of the E864 experiment. The model
essentially fails to describe the ratio.  For larger impact
parameters, however, we cannot conclude with certainty that the
hadronic mechanisms are inadequate in describing the data. While the
data drops dramatically as the impact parameter increases, the calculation
remains relatively flatter, and even rises somewhat for an energy 
independent ${\bar \Lambda}$ annihilation cross--section.

In conclusion, the large ${\bar \Lambda}/\bar p$ ratios in central
Au+Au collisions at $11.6~{\rm A\cdot GeV}/c$ are not easily explained
by hadronic mechanisms.  This is no longer true at larger impact
parameters, and it is interesting to speculate if QGP formation in
central events might be a possible explanation. Of course, more accurate
measurements of the $\bar \Lambda$ annihilation cross--section in the
relevant energy range are needed, as well as a better
understanding of hadronic medium effects in ${\bar \Lambda}$ and
${\bar p}$ production and annihilation, before any more definite
conclusion can be reached.

Acknowledgments: Gerd Welke acknowledges useful discussions with
Pawel Danielewicz, J\"org Aichelin and Wolfgang Cassing.  GW and
G.~Wang acknowledge the hospitality of the Institute for Nuclear
Theory, Seattle, where this work was completed.  This work was
supported in part by the U.S. Department of Energy under Grant No.
DE-FG02-93ER40713.

\newpage

\section{TABLES}

\noindent TABLE~1. Thermal and experimental particle ratios
for central Au+Au collisions at $11.6~{\rm GeV}/c$.  The parameter
ranges are $T_0= 120\pm14~{\rm MeV}$, $\mu_b=556\pm19~{\rm MeV}$,
$\mu_s= 111\pm14~{\rm MeV}$, and $\mu_e=-14\pm2~{\rm MeV}$.  The
second column lists the range of the various constraint ratios considered.
\vskip 0.3in
\begin{center}
\begin{tabular}{|c|c|c|ccc|}
\hline 
 & Constraint & & 
\multicolumn{3}{c|}{Data} \\ \cline{4-6}
Ratio & Range & Thermal Model & Ratio & Rapidity & Ref. \\ [0.3mm] \hline
$K^+$/$\pi^+$ & 0.16--0.28 & $0.23\pm 0.03$ & $0.22\pm 0.01$ & 
0.5--1.3 & \cite{Gonin} \\ [0.3mm]
$K^+/K^-$ & 4.0--6.0   & $4.73\pm 0.53$ & $5.0\pm 1.0$ &
0.5--1.3 & \cite{Gonin} \\ [0.3mm]
$K^-$/$\pi^-$& -- & $(3.50\pm 0.62)\times 10^{-2}$
        & 0.028 & 1.2--2.0 &\cite{Hiroyuki} \\ [0.3mm]
$\pi^+/p$ & 0.6--1.2   & $0.71\pm 0.09$& -- & -- & -- \\ [0.3mm]
$\pi^-/p$ & 0.8--1.4 & $1.00\pm 0.10$ & 1.00 & 1.2--2.0 &
\cite{Hiroyuki} \\ [0.3mm]
$\Lambda/p$ & -- & $0.16\pm 0.02$& -- & -- & -- \\ [0.3mm]
$\bar p/p$ & -- & $(3.48 \pm  3.44) \times 10^{-4}$ & 
$(5 \pm  2) \times 10^{-4}$ & 1.6~($p_\perp\!\sim\!0$) & 
\cite{Claude} \\ [0.3mm]
$\bar{\Lambda}/\bar{p}$ & -- & $1.58\pm 0.30$& 3.5 & 
1.6~($p_\perp\!\sim\!0$) 
& \cite{e864b} \\ [0.3mm]
 &  & & $> 2.3$ (98\% CL) 
& 1.6~($p_\perp\!\sim\!0$) & \cite{e864b} \\ [0.3mm] \hline
\end{tabular}
\end{center}

\vskip 0.9in

\noindent TABLE 2 Survival probabilities $P_s$, the ratio of surviving
converted
${\bar \Lambda}^c$'s to surviving ${\bar p}^s$,
and final ${\bar\Lambda}/{\bar p}$ ratio for various $\alpha$ and
$\tau=0~{\rm fm}/c$, in central (${\buildrel < \over \sim}10\%$) Au+Au
cascade simulations.  
\vskip 0.25in
\begin{center}
\begin{tabular}{|c|ccc|c|} \hline
   \raisebox{-0.1cm}{$\alpha$} & \raisebox{-0.1cm}{$P_{s}({\bar\Lambda})$
    (\%)} & \raisebox{-0.1cm}{$P_{s}({\bar p})$ (\%)} &
    \raisebox{-0.1cm}{${\bar\Lambda}^{c}/{\bar p}^s$}&
    \raisebox{-0.1cm}{${\bar\Lambda}/{\bar p}$} \\ [0.3mm]
    \hline 0.0 & $12.1\pm 0.3$ & $1.5\pm
    0.1$ & $0.75\pm0.07$ & $2.4\!+\!0.2\!-\!1.0$ \\ [0.3mm]
0.2 & $6.0\pm 0.2$ &
    $1.5\pm0.1$ & $0.45\pm0.06$ & $1.2\!+\!0.1\!-\!0.5$ \\ [0.3mm]
0.4 & $3.3\pm0.2$ &
    $1.5\pm 0.1$ & $0.38\pm0.06$ & $0.7\!+\!0.1\!-\!0.3$ \\ [0.3mm]
0.7 & $1.5\pm0.05$ &
    $1.5\pm0.1$ & $0.16\pm0.02$ & $0.3\!+\!0.04\!-\!0.1$ \\ [0.3mm] \hline
\end{tabular}
\end{center}

\newpage

\section{FIGURE CAPTIONS}

\noindent FIGURE~1.~ ~(a) 
Thermal ${\bar \Lambda}/{\bar p}$ ratios as a function of
$K^{+}/K^{-}$ for various freeze--out temperatures $T_0$.  (b) Thermal
${\bar \Lambda}/{\bar p}$ ratios as a function of $K^+/\pi^+$.  The
error bars/scatter points indicate values consistent with constraints
not shown (see Table~1 for details). The solid lines are to guide the
eye only; the dashed line represents the 98\% lower CL of the E864
measurements.

\vskip 0.4in

\noindent FIGURE~2.~ ~The annihilation cross--sections of
$\bar{p}$ ($\bar{\Lambda}$) with nucleons, as a function
of the incident momentum of the $\bar{p}$ ($\bar{\Lambda}$). The solid 
and dashed lines are fits to the ${\bar p}$ and $\bar{\Lambda}$ data, 
respectively, while the dot--dashed curve is the extreme fit to the ${\bar 
\Lambda}$ data (with $\alpha=0$ in Eq.~(\ref{PLbar_low})). 
The $\bar{\Lambda}$ data (diamonds) are 
from Ref.~\cite{eisele}.

\vskip 0.4in

\noindent FIGURE~3.~ ~${\bar \Lambda}/{\bar p}$ ratios calculated in
the cascade model, for various values of $\alpha$, and $\tau$ for central
Si+Pb and Au+Au collisions. The solid lines are results of a geometric model
calculation of Ref.~\cite{GJWang}, while the dashed lines show the
E864 lower 98\% CL \cite{e864b}.

\vskip 0.4in

\noindent FIGURE~4.~ ~${\bar \Lambda}/{\bar p}$ ratios as a function of
centrality for Au+Au and zero formation time $\tau$.  All lines are
drawn to guide the eye only. The upper solid line shows the E864 most
probable values, the lower solid line represents the E864 98\% CL
\cite{e864b,QM97E864}.  The crosses are cascade calculations for $\alpha=0$,
while the cross--box points are for $\alpha=0.4$. Some points are
offset slightly to clarify the error bars.

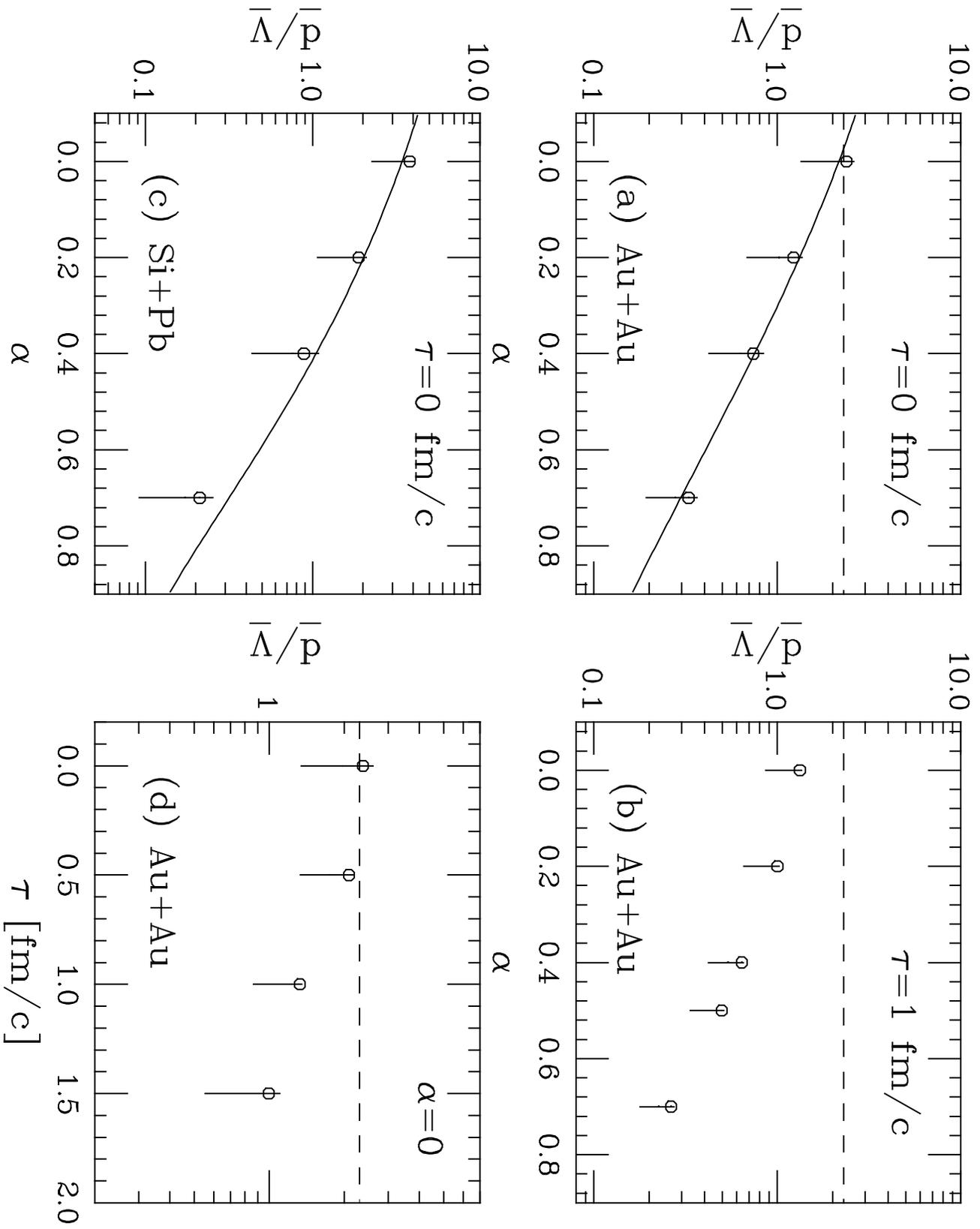
\end{document}